\documentclass[12pt]{article}

\advance\voffset by -2.0cm \advance\hoffset by -1.25cm
\usepackage{amsmath}
\usepackage{amssymb}
\usepackage{amsthm}
\usepackage{amsfonts}

\theoremstyle{definition}

\newcommand{\RR}{\mathbb{R}} 
\newcommand{\ZZ}{\mathbb{Z}} 

\DeclareMathOperator{\tr}{tr} 

\def\l{\label}
\def\S{{\cal S}}
\def\U{{\cal U}}

\def\s{{\tt s}}

\newcommand{\be}{\begin{equation}}
\newcommand{\ee}{\end{equation}}

\newlength{\oldcolsep}

\begin{document}

\title{``Thermodynamique cach\'ee des particules'' and the quantum potential}

\author{Marco Matone}\date{}

\maketitle

\begin{center}Dipartimento di Fisica ``G. Galilei'' and Istituto
Nazionale di Fisica Nucleare \\
Universit\`a di Padova, Via Marzolo, 8 -- 35131 Padova,
Italy \\
e-mail: matone@pd.infn.it\end{center}

\bigskip

\bigskip

\begin{abstract}
\noindent According to de Broglie, temperature plays a basic r\^ole in quantum Hamilton-Jacobi theory.
Here we show that a possible dependence on the temperature of the integration constants of the relativistic quantum Hamilton-Jacobi
may lead to corrections to the dispersion relations.
The change of the relativistic equations is simply described by means of a thermal coordinate.

\bigskip



\end{abstract}

\newpage

In Louis de Broglie view temperature plays a central r\^ole in quantum Hamilton-Jacobi theory.
His investigation is summarized in the beautiful paper \cite{deBroglie}.
A related issue concerns possible deformations of the dispersion relations in the framework of
the relativistic quantum Hamilton-Jacobi equation considered in \cite{ShawnLane}. The quantum versions of the Hamilton-Jacobi equations have been
derived by first principles in \cite{1,BFM,2}. Interestingly, such a formulation is
strictly related to classical-quantum  \cite{Faraggi:1996rn} and Legendre
\cite{Legendre} dualities.

Consider the stationary Klein-Gordon equation
$(-\hbar^2c^2\Delta+m^2c^4-E^2)\psi=0$.
In one-dimensional space the associated quantum Hamilton-Jacobi
equation is \be (\partial_q S_0)^2+m^2c^2-{E^2\over
c^2}+{\hbar^2\over2}\{S_0,q\}=0 \ , \label{unodim}\ee where
$\{f,q\}={f'''\over f'}-{3\over2}\big({f''\over f'}\big)^2$ denotes
the Schwarzian derivative of $f$. The term
$Q={\hbar^2\over4m}\{S_0,q\}$,
is the quantum potential and
$p=\partial_q \S_0$
the conjugate momentum. It is crucial that $Q$, which is always non-trivial \cite{1,BFM,2}, is quite different with respect to the one by de Broglie and Bohm.
In \cite{ShawnLane} it has been shown that, upon averaging the mean speed on the period $[q, q+{\pi\over k}]$, $k=\sqrt{E^2-m^2c^4}/\hbar c$, one gets
$v= c{\sqrt{E^2-m^2c^4}\over E}/L_1$
with $L_1\leq 1$ an integration constant of (\ref{unodim}). Here we investigate
a possible dependence of the initial conditions of (\ref{unodim}) on the
temperature $T$. More precisely, we will consider some analogies with thermal field theories
suggesting a correction for the relativistic speed depending on the ratio $m/T$
\be
v=c{\sqrt{E^2-m^2c^4}\over E}\alpha(m/T) \ .
\label{ansatz}\ee

Since the quantum potential plays the central r\^ole in our
construction, it is worth stressing that the derivation and
formulation of the quantum version of the Hamilton-Jacobi equations
of \cite{1,BFM} is quite different from the one by de Broglie-Bohm.
In particular, the quantum potential is basically different. Let us
consider the case of non-relativistic quantum mechanics. de
Broglie-Bohm made the identification $\psi=Re^{{i\over\hbar} \S_0}$
with $\psi$ the wave-function, not just a general complex solution
of the Schr\"odinger equation, as it should be on mathematical
grounds \cite{1,BFM}. This leads to the following trouble. Consider
the case in which the wave-function is the eigenfunction of the
Hamiltonian corresponding to a one-dimensional bound state, {\it
e.g.} the harmonic oscillator. In this case $\psi$ is proportional
to a real function. It follows that $\S_0$ is a constant, so that
the conjugate momentum is trivial. This would imply that at the
quantum level the particle is at rest and starts moving in the
classical limit. This is a well-known paradox observed by Einstein.
In the derivation considered in \cite{1,BFM} such a paradox is
naturally resolved by construction: $\{\S_0,q\}$ is well-defined if
$\partial_q \S_0$ is never vanishing. In particular, the general
solution of the non-relativistic quantum stationary Hamilton-Jacobi
equation, which is formally equivalent to (\ref{unodim}), with
$(m^2c^4-E^2)/2mc^2$ replaced by $V-E$, is \be
e^{{2i\over\hbar}\S_0\{\delta\}}=e^{i\alpha}{w+i\bar\ell\over
w-i\ell} \ , \l{s}\ee with $w=\psi^D/\psi\in\RR$, where now $\psi$
and $\psi^D$ are two real linearly independent solutions of the
stationary Schr\"odinger equation. $\delta=\{\alpha,\ell\}$,
$\alpha\in\RR$ and $\ell=\ell_1+i\ell_2$ are integration constants.
Note that $\ell_1\ne 0$ even when $E=0$, equivalent to having
$\S_0\ne cnst$, which is a necessary condition to define
$\{\S_0,q\}$. This implies a non-trivial $\S_0$, even for a particle
classically at rest. One has
$\psi={\S_0'}^{-1/2}\big(Ae^{{i\over\hbar}\S_0}+Be^{-{i\over\hbar}\S_0}\big)$,
 and $\psi\in\RR$ implies $|A|=|B|$, so that $\S_0$ is
non-trivial. The $\hbar\to0$ limit is subtle and leads to the
appearance of fundamental constants \cite{1,BFM,2}. A basic
observation in \cite{ShawnLane} is that averaging the period of the
oscillating terms, which also appear in the non-relativistic case,
leads to $\hbar$-independent solutions that, besides including the
standard one, describe other solutions depending on the values of
the integration constants (see also later). We note that the
above decomposition of $\psi$, now called bipolar decomposition, is
successfully used in studying molecular trajectories.

Another distinguished feature of the formulation introduced in
\cite{1,BFM} is that energy quantization follows without any
axiomatic interpretation of the wave-function.
As a result, the quantum potential, intrinsically  different from
the one by de Broglie-Bohm, is always non-trivial. Like $mc^2$, it
plays the r\^ole of intrinsic energy and it is at the basis of the
quantum behavior. For example, besides energy quantization, it makes
transparent its r\^ole in the tunnel's effect, where guarantees that
the conjugate momentum always takes real values.

The general solution of (\ref{unodim}) is (\ref{s}), where now $\psi$ and $\psi^D$
are two real linearly independent solutions of the
Klein-Gordon equation.
Following Floyd \cite{Floyd}, time parametrization is defined by Jacobi's theorem
$t={\partial {\S_0}\over\partial E}$.
Since $p=\partial_q \S_0$, it gives the group velocity $v=\partial
E/\partial p$. Set $L=k\ell$, $L_1=\Re L$, $L_2=\Im L$. In the case $|L|=1$ the mean
speed reads
$$ v=c{\sqrt{E^2-m^2c^4}\over E}{1+L_2
\, \sin(2kq)\over L_1} \ . $$ The integration
constants $L_1$ and $L_2$ may depend on particle's quantum numbers,
energy and fundamental constants as well. As we said, here we will
consider an intriguing dependence of $L$ on the temperature. Such a
possibility is related to a new way of considering the $\hbar\to 0$
limit which has been introduced in \cite{ShawnLane}. Let us first
note that because of the $\hbar^{-1}$ term in $\sin(2kq)$, that
typically is very strongly oscillating, the $\hbar\to 0$ limit is
not well-defined.
 One possibility, considered in \cite{1,BFM}, is
that the integration constants may depend on $E$, $\hbar$ and other
fundamental constants. In this way the term $\hbar^{-1}$ is cured by
a suitable dependence on $\hbar$ of $L$, a procedure that leads to
consider the Planck length (note that $\ell$ has the dimension of a
length).
The new way of getting the classical limit considered in
\cite{ShawnLane} is based on the averaging of the oscillations.
In particular
$$ \langle v\rangle ={k\over \pi}\int^{q+{\pi\over
k}}_qv(q')dq'= c{\sqrt{E^2-m^2c^4}\over E}{1\over L_1} \ ,
$$
which is reminiscent of the Dirac's averaging of the oscillating part of the
free electron's speed ${i\over2}c\hbar {\dot\alpha_1^0}e^{-2iHt/\hbar}H^{-1}$ (see Dirac's treatment
of the free electron in his book).

Note that a possible dependence of $L$ on the temperature should
not be a surprise. Actually, as observed by Dirac in computing the
speed of the free electron, the uncertainty principle implies that
it does not make sense considering scales which are much shorter
than the Compton wavelength. On other hand, averaging on the period
may lead to the breaking of the Lorentz group.
Furthermore, it is natural to expect that such an averaging leads to
a dependence on the degrees of freedom of the averaged space domain,
which in general is not the empty space. In doing this we should
remind the analogy with spontaneous symmetry breaking of Lorentz
group in thermal QFT, we will shortly discuss later. Also note that
the quantum potential is strictly related to the Fisher information
and Shannon entropy \cite{Reginatto}. It
is then natural to expect that also here the temperature plays the
central r\^ole.

Remarkably, dependence of $L$ on the temperature is allowed just as
a consequence of a basic physical property such as linearity of
quantum mechanics. In the case of the Klein-Gordon equation, as in
the case of the Schr\"odinger equation, such a property is
equivalent to the invariance of the Schwarzian derivative under
M\"obius transformations. In this respect, note that a basic
identity at the basis of \cite{1,BFM,2} is that $(\partial_q\S_0)^2$
can be expressed as the difference of two Schwarzian derivatives
$\left({\partial\S_0\over\partial q}\right)^2=
{\beta^2\over2}\left(\{e^{{2i\over\beta}\S_0},q\}-\{\S_0,q\}\right)$,
that forces us introducing the dimensional constant $\beta$, that is
$\hbar$. We also note the appearance of the imaginary factor. It
follows that Eq.(\ref{unodim}) is equivalent to the Schwarzian
equation
$\{e^{{2i\over\hbar}\S_0},q\}= {2\over \hbar^2c^2}(E^2-m^2c^4)$,
which is invariant under a M\"obius transformation of $e^{{2i\over\hbar}\S_0}$
so that we can also assume
that $\alpha$ and $\ell$ depend on the temperature
so that Eq.(\ref{s}) has the form \be
e^{{2i\over\hbar}\S_0\{\delta(T)\}}=e^{i\alpha(T)}{w+i\bar\ell(T)\over
w-i\ell(T)} \ . \l{sdue}\ee This provides an intriguing possibility
in the case of non-relativistic quantum mechanics. Actually,
possible dependence on the temperature provides a sort of dynamics
non detected by the wave-function and may open a new view on its
collapse. As the temperature changes,
$e^{{2i\over\hbar}\S_0\{\delta(T)\}}$ moves on the boundary of the
Poincar\'e disk (discrete jumps by M\"obius transformations are also
allowed) with $\{e^{{2i\over\hbar}\S_0},q\}$ remaining invariant.
This happens also when $\{e^{{2i\over\hbar}\S_0},q\}$ is considered
in the case of the relativistic quantum Hamilton-Jacobi equation
(\ref{unodim}). In particular, $\{e^{{2i\over\hbar}\S_0},q\}={2\over
\hbar^2c^2}(E^2-m^2c^4)$ is invariant under variations of both the
coordinate and the temperature.

Eq.(\ref{sdue}) may be related to $t-T$ duality and non-commutative
geometry (see \cite{Matone:2006mw} and references therein). In this
respect, it is worth recalling de Broglie view \cite{deBroglie}. In
particular, in the framework of relativistic thermodynamics, he
considered $t_m=h/mc^2$ as a particle internal time and identified
it with an internal temperature $KT_m=mc^2$.
de Broglie also suggested that, in the case of non-relativistic
quantum mechanics, the collapse of the wave-function is related to a
sort of particle's Brownian interaction with the environment.

Time-temperature duality is a basic feature of QFT, related to the analogy with
classical statistical mechanics where the inverse temperature plays
the r\^ole of imaginary time. Time and temperature also mix in
performing analytic continuation along complex paths in the
path-integral.
Another well-known analogy concerns the transition
amplitude for a particle for the time $it$ that coincides with the
classical partition function for a string of length $t$ at
$\beta=1/\hbar$ \cite{Polyakov}.
 It is widely believed that such a dualities are deeply related to
the properties of space-time and should emerge in the string
context.

In \cite{Matone:2006mw} it has been considered a string model where
time-temperature and classical-quantum dualities, emerge naturally.
It turns out that the solitonic sector of compactified strings have
a dual description as quantum statistical partition function on
higher dimensional spaces, built in terms of the Jacobian torus of
the string worldsheet and of the compactified space. More precisely,
in \cite{Matone:2006mw} it has been shown that in the case of
compactification on a circle \be\sum_{m,n\in\ZZ^g}e^{-\beta
S_{m,n}}=\tr\, e^{-\beta H}\ , \l{laprima}\ee where
$\beta=2R^2/\alpha'$, with $R$ the compactification radius, and the
Hamiltonian $H$ is $\Delta_{J_\Omega}/2\pi$, with
$\Delta_{J_\Omega}$ the Laplacian on the Jacobian torus $J_\Omega$
of the worldsheet. Eq.(\ref{laprima}) is just the direct consequence
of the stronger identity
$H\Psi_{m,n}=S_{m,n}\Psi_{m,n}$,
that is the set $\{
S_{m,n}|m,n\in\ZZ^g\}$ {\it coincides} with the spectrum of $H$, a result deeply related to the
theory of Riemann surfaces \cite{Riemannsurfaces}.

Temperature-time duality naturally emerges as a
consequence of the complexified version of $T$-duality, a
fundamental feature of string theory. By (\ref{laprima}) the standard
$T$-duality corresponds to the invariance, up to a multiplicative
term given by powers of $\beta$, of the partition function under inversion of
the temperature
$\beta\longrightarrow {1\over\beta}$.
Complexification of $\beta$ has basic motivations which interplay
between physics and geometry. Set
$\omega(A)=\tr\, \big(A
e^{-\beta H}\big)/\tr\, e^{-\beta H}$. Using the invariance of the
trace under cyclic permutations we have
\be
\omega(A(t)B)=\omega(BA(t+i\beta))\ .
\l{diKMS}\ee
In such a
context the complexification of $\beta$ naturally appears in
globally conformal invariant QFT \cite{NikolovYG}. It is worth noticing that both in \cite{NikolovYG} and in the BC
system \cite{BC}
the KMS (Kubo-Martin-Schwinger) states \cite{NONC} play a central r\^ole. In particular, in the limit of
$0$-temperature the KMS states may be used to define the concept of
point in noncommutative space, a basic issue in string theory \cite{noncommutative}.

A nice feature of the possible temperature dependence of the integration constants of (\ref{unodim}) is that it does not arise as a consequence of a specific
interaction, rather, as shown above, it is just due to the linearity of the Klein-Gordon equation, or, equivalently, of the M\"obius symmetry of
the Schwarzian derivative.

The above remarks suggest asking whether there exists an $L(T)$ such that the resulting expression for the velocity
agrees with the experimental data. Let us then go back to the KMS states and note that the Fourier transformed form of (\ref{diKMS})
\cite{OjimaIS}
\be
FT[\omega(AB)](p)\equiv \int d^d x e^{ip(x-y)}\omega(A(x)B(y))=e^{-\beta p_0}FT[\omega(BA)](-p) \ ,
\l{rt}\ee
shows that $FT[\omega(AB)](p)$ and $e^{-\beta p_0}FT[\omega(BA)](-p)$ cannot be both Lorentz invariant.
Spontaneous breakdown of Lorentz boost symmetry at $T\neq0$ is nicely seen considering the $(1,1)$ component of the $2\times2$ matrix propagator
of a scalar field that at the tree level is \cite{OjimaIS}
$$
FT[\omega(\varphi\varphi)](p)={1\over p^2-m^2+i\epsilon}+2\pi\delta(p^2-m^2){e^{-\beta|p_0|}\over 1-e^{-\beta|p_0|}} \ ,
$$
which breaks Lorentz
$\omega([M_{01},T\varphi(x)\varphi(y)])\neq0$.

Even if we are considering a breaking of the Lorentz group from the quantum potential, which is a highly nonlocal term, it is clear that
the analogy with thermal QFT suggests considering in our approach
$e^{-{mc^2\over KT}}=e^{-{T_m\over T}}$,
as order parameter of the Lorentz symmetry breaking. The crucial
property of the KMS states is (\ref{diKMS}), or equivalently
(\ref{rt}), showing that time evolution is invariant, upon
commutation, under an imaginary shift of the time which is inversely
proportional to the temperature. In the case of the quantum
Hamilton-Jacobi equation, time is generated by Jacobi's
theorem. Then one gets $t$ as a function of $q$ and a shift of time
corresponds to a shift of $q$. On the other hand, this corresponds
to a jump of $e^{{2i\over\hbar}\S_0}$ on the boundary of the
Poincar\'e disk, and may be compensated by changing $L(T)$ that, depending on its functional
structure, may correspond to a change of $T$.

In the present formulation the dependence on $m/T$ as
order parameter of the breaking of Lorentz symmetry
comes out as an effect
due to the quantum potential and it is not seen as due to some
particle interaction. Remarkably, even
in this case there is an analogy with the thermal breakdown of
symmetries in thermal field theory since the spontaneous breakdown
of symmetries occurs even in cases with no interactions \cite{UmezawaYQ}.

Let us now show that the modified relativistic dispersion relation induced by
(\ref{ansatz}) is
\be E^2= p^2c^2\alpha^2(m/T)+m^2c^4 \ .
\l{dispepsiolo}\ee
Replacing $(m^2c^4-E^2)/2mc^2$ by $V-E$,
(\ref{unodim}) is formally equivalent to the non-relativistic
quantum stationary Hamilton-Jacobi equation. It follows that, as
observed in
 \cite{1,BFM},  Eq.(\ref{unodim}) can be seen as a deformation of the classical
relativistic Hamilton-Jacobi equation by a ``conformal factor''. Actually, noting that
$
\{\S_0,q\}=-(\partial_q\S_0)^2\{q,\s\}$,
$\s=\S_0(q)$, we see that (\ref{unodim}) is equivalent to
\be
E^2=\left({\partial\S_0\over\partial q}\right)^2c^2\left[1-\hbar^2\U(\S_0)\right]+m^2c^4 \ ,
\l{hsxdgyij}\ee
with $\U(\S_0)=\{q,\s\}/2$, the canonical potential introduced in
the framework of $p$--$q$ duality \cite{1,BFM}.  Eq.(\ref{hsxdgyij}) can be
expressed in the form
$E^2=({\partial_{\hat q}\S_0})^2c^2+m^2c^4$
where
$d\hat q={dq/\sqrt{1-{\hat\beta}^2(q)}}$
with ${\hat\beta}^2(q)=\hbar^2\{q,\s\}/2$. Integrating
 \be \hat
q=c\int^qdq'{\partial_{q'}\S_0\over\sqrt{E^2-m^2c^4}}=
c\int^{\S_0(q)}{d\s\over\sqrt{E^2-m^2c^4}} \ . \l{questa}\ee Comparison with
the dispersion relation (\ref{dispepsiolo}) shows that the ansatz (\ref{ansatz}) is equivalent to the thermal deformation of the coordinate
$$
\lim_{``\hbar\longrightarrow0"}\hat q=q_T={q\over\alpha(m/T)} \ ,$$
so that the relativistic quantum Hamilton-Jacobi equation (\ref{unodim}) becomes
$$E^2=\left({\partial\S_0\over\partial q_T}\right)^2c^2+m^2c^4 \ ,$$
which is Eq.(\ref{dispepsiolo}).
Eq.(\ref{questa}) is related to the relativistic classical action
whose modification,
leading to (\ref{dispepsiolo}), is $$ L=-{mc^2\over\gamma_T} \ ,
$$ with
$\gamma_T=1/\sqrt{1-{v^2\over c^2\alpha^2(m/T)}}$,
we have
$$
p={\partial L\over\partial v}={mv\gamma_T\over \alpha^2(m/T)} \ ,
$$
and
$$ E=pv-L=mc^2\gamma_T \ ,
$$
so that we get (\ref{dispepsiolo}) and by $v=\partial_p E$,
$$v={pc^2\over E}\alpha^2(m/T) \ ,$$
which is, by (\ref{dispepsiolo}), Eq.(\ref{ansatz}).

\vspace{.333cm}

\newpage

\bibliography{apssamp}

\end{document}